\documentclass[%
 reprint,
 amsmath,amssymb,
 aps,
]{revtex4-2}

\usepackage{graphicx}% Include figure files
\usepackage{dcolumn}% Align table columns on decimal point
\usepackage{bm}% bold math
\begin{document}

\preprint{APS/123-QED}

\title{A Joint-Chirp-Rate-Time-Frequency Transform \\for BBH Merger Gravitational Wave Signal Detection}% Force line breaks with \\

\author{Xiyuan Li}
 \email{xli2522@uwo.ca}
\affiliation{Department of Physics and Astronomy, University of Western Ontario, London, Ontario, Canada} %Lines break automatically or can be forced with \\

\author{Martin Houde}
 \email{mhoude2@uwo.ca}
\affiliation{Department of Physics and Astronomy, University of Western Ontario, London, Ontario, Canada}

\author{Jignesh Mohanty}
 \email{mohantyj@iitk.ac.in}
\affiliation{%
 Department of Physics, Indian Institute of Technology Kanpur, India
}%

\author{S. R. Valluri}
 \email{valluri@uwo.ca}
\affiliation{
 Department of Physics and Astronomy, University of Western Ontario, London, Ontario, Canada
}%
\altaffiliation[Also at]{
 Mathematics, Kings University College, University of Western Ontario, London, Ontario, Canada
}%
\date{March 4, 2023}% It is always \today, today,

\begin{abstract}
Low-latency detection of Binary Black Hole (BBH) and Binary Neutron Star (BNS) merger Gravitational Wave (GW) signals is essential for enabling multi-messenger observations of such systems. The merger GW signals have varying frequencies and are contaminated by non-stationary noises. Earlier studies of non-templated merger signal detection techniques used traditional Fourier transform-based time-frequency decomposition methods for spectrogram generation, which have had difficulties identifying rapid frequency changes in merger signals with heavy background noise. To address this problem, we introduce the Joint-Chirp-rate-Time-Frequency Transform (JCTFT), in which complex-valued window functions are used to modulate the amplitude, frequency, and phase of the input signal. In addition, we outline the techniques for generating chirp-rate-enhanced time-frequency spectrograms from the results of a JCTFT. We demonstrate an average of 14\% improved merger detectability among simulated detector signals with Signal-to-Noise Ratios between 6 and 10 using the InceptionV3 image classification neural network compared to the same network trained with Q-transform spectrograms. The JCTFT is a general transformation technique that can be applied to existing and third-generation GW detector signals. Further studies will aim to improve the efficiency and performance of the JCTFT.
\end{abstract}

\maketitle

%\tableofcontents

\section{Introduction}
The direct detection of gravitational waves (GW) in 2015 by the Laser Interferometer Gravitational-Wave Observatory (LIGO) and Virgo (European Gravitational Wave Observatory) scientific collaboration opened up a new era of GW research and provided researchers with never-before-seen data that allowed insights into the cosmic events that produce such signals\cite{PhysRevLett.116.061102}. Some of the most prominent sources of GW signals are Binary Black-Hole (BBH) and Binary Neutron Star (BNS) mergers\cite{PhysRevLett.116.241103, PhysRevLett.119.161101}. With the addition of the Japanese KAGRA detector, formerly known as the Large Scale Cryogenic Gravitational Wave Telescope, and LIGO India, next-generation ground and space-based detectors such as the Einstein Telescope, Cosmic Explorer, and the Laser Interferometer Space Antenna\cite{KAGRA:2020tym, Saleem_2022, 2020JCAP, 2019BAAS...51g..35R,  2017arXiv170200786A} underway, developing tools that allow for the fast and accurate identification of GW signals from merger events will allow immediate coordination between observatories to capture a fuller picture of the process leading up to the merger\cite{george_huerta_2018}. \\
There are two main approaches for merger signal detection: templated and non-templated searches\cite{PhysRevD.85.122006, drago_klimenko_lazzaro_2021}. Currently, one of the most reliable methods is matched filtering using simulated merger waveforms. This templated search technique matches the detector signal to a collection of waveform templates in the template bank, and outputs the best match based on a series of criteria. In the context of BBH and BNS merger GW signals, simulated merger waveforms obtained using a combination of numerical techniques are correlated with detector signals, thereby producing matched filters\cite{ PhysRevD.54.7108, owen_sathyaprakash_1999, doi.org/10.1007/BF02705194, PhysRevD.71.062001}. A high presence of the template waveform in the detector signal produces a good match. The matched waveform is then considered as a detected signal if it passes a series of vetoing algorithms\cite{Nitz:2017svb, PhysRevD.90.082004, Abbott_2018}. \\
Many researchers have also attempted non-templated search techniques based on time-frequency spectrograms, where the algorithm looks for sudden appearances of chirp signals that resemble the characteristics of a BBH or BNS merger. The fundamental technique used to generate spectrograms, such as the Fourier Transform (FT), can be loosely considered as a type of matched filtering. Because FT-inspired spectrogram generation techniques contain no actual merger waveforms and the filter bank is rather simple, we excluded the use of simple spectrograms from templated searches. An example of such an algorithm is coherence waveBurst, which has been applied to merger signal searches and analysis since the early days of BBH merger detection\cite{drago_klimenko_lazzaro_2021, Klimenko_2016}. The performance of such spectrogram-based, non-templated searches may vary depending on the time-frequency decomposition method. The short-time-Fourier-transform (STFT), Q transform (QT), and other modified wavelet transform methods such as the Wilson-Daubechies transform are commonly used\cite{wei_khan_huerta_huang_tian_2021, gwpy, drago_klimenko_lazzaro_2021}.  \\
Non-templated detection techniques based on time-frequency decomposition are gaining traction again owing to recent advances in machine learning algorithms\cite{_lvares_2021}. These algorithms are extremely effective in identifying underlying connections within complicated input data and predicting outcomes based on efficient statistical procedures\cite{arxiv.1502.03167, arxiv.1506.02142}. Examples that use spectrograms as an integral part of a classification ANN include the Gravity Spy Project, where GW detector glitches are classified using an image ANN\cite{zevin_coughlin_bahaadini_2017}, and the merger GW signal forecasting ANN built on the pre-trained image classification network ResNet-50\cite{wei_huerta_yun_2021}. However, there are several shortcomings associated with the use of traditional time-frequency decomposition methods for GW signal detection and classification. For instance, the risk of degrading weak signals in map-whitening, the inability to distinguish low-frequency and low-power inspiral signals from background noise\cite{drago_klimenko_lazzaro_2021}, and the difficulty in capturing the rapid frequency change of a merger. Weaker signals make accurate detection even more difficult to achieve. Although many of the problems can be addressed through better machine learning architectures, high-performance hardware, and more training data, a signal decomposition method that provides better pre-processed input data has the potential to produce immediate improvements in detection sensitivity and accuracy, particularly for low Signal-to-Noise Ratio (SNR) merger GW signals. \\
In this paper, we introduce the complex window-based Joint-Chirp-rate-Time-Frequency Transform (JCTFT) for non-templated merger chirp signal detection. JCTFT uses a frequency-related complex window function to vary its time-sampling interval and the degree of frequency shift for different frequency components. Lower frequency signals were sampled for longer periods of time to account for their longer durations. In comparison, higher-frequency signals were sampled for shorter periods of time to capture the quick frequency changes in a chirp signal. This method is distinct from existing windowed STFT methods and most commonly used wavelet transforms. The complex window function contains terms that do not satisfy the wavelet admissible condition and the transform produces results in three dimensions instead of only time and frequency. JCTFT introduces frequency shifts represented by non-linear time scales that are not present in traditional time-frequency decomposition methods, thereby generating more prominent frequency peaks of chirp signals. Some studies have discussed intents or techniques reminiscent of JCTFT\cite{PhysRevD.62.122001,alkishriwo_chaparro_2012, stockwell_mansinha_lowe_1996, ComplexDaubechies1997, pinnegar_mansinha_2004}, in particular, the chirplet transform\cite{482123}. However, we find our definition is unique in terms of the complex-window design, window application procedure, result processing, and intent for application. We found an average of 14\% improved merger detection accuracy or lower false positive rate in two InceptionV3 convolutional neural networks trained separately with JCTFT and QT spectrograms.\\
The remainder of this paper is organized as follows. In Section two, we introduce the idea behind the linear chirp approximation of BBH merger GW signals, the JCTFT, and the technique for generating chirp-rate-enhanced time-frequency spectrograms. Section three provides the processed three-dimensional JCTFT results, spectrograms of high and low-SNR simulated BBH merger GW signals, and the ANN classification statistics. In Section four we discuss the significance, applications, and potential improvements associated with the JCTFT. Section five presents the conclusions and outlines the broader applications of the complex window function.
%%%%%%%%%%%%%%%%%%%%%%%%Linear Chirp Base for BBH Merger GW Signal Representation%%%%%%%%%%%%%%%%%%%%%%
\section{Methods}
\subsection{Linear Chirp Base for BBH Merger GW Signal Representation}
A simplified BBH merger GW instantaneous frequency (IF) model for lower frequency, near-circular orbit systems, is given by\cite{ligo_virgo_2016}
\begin{equation}
{IF_{merger}}^{-8/3}(t)=\frac{(8\pi)^{8/3}}{5}(\frac{GM_{chirp}}{c^3})^{5/3}(t_c-t),
\label{equ: IF merger ^(-8/3)}
\end{equation}
following a 2PN correction. Then 
\begin{equation}
{IF_{merger}}(t)=\frac{5^{3/8}}{8\pi}(\frac{c^3}{GM_{chirp}})^{5/8}(t_c-t)^{-3/8},
\label{equ: IF merger}
\end{equation}
where $G$ is the gravitational constant, $t_c$ is the coalescence time, and $M_{chirp}$ is the chirp mass of the BBH system defined by the individual masses $m_1$ and $m_2$ via
\begin{equation}
M_{chirp} = \frac{(m_1m_2)^{3/5}}{(m_1+m_2)^{1/5}}; \; m_1 > 0, \; m_2 > 0.
\end{equation}
A complex linear chirp signal can be represented as
\begin{equation}
h^{chirp}_{linear}(t) = Ae^{i2\pi(\Omega t+\gamma t^2)},
\end{equation}
where $A$ is the amplitude, $\Omega$ is the starting frequency, and $\gamma$ is the chirp rate. A simple differentiation of the power term $\Omega t+\gamma t^2$ yields the IF
\begin{equation}
IF^{chirp}_{linear}(t, \Omega) = 2\gamma t+\Omega.
\label{equ:linear chirp IF}
\end{equation}
It is easy to show that the merger frequency model $IF_{merger}(t)$ is an entire function on $t \in (0,t_c)$, and its chirp rate $IF^{'}_{merger}(t)$ is monotonic. By appropriate choices of parameter values, there always exists a linearly changing frequency in the form of equation (\ref{equ:linear chirp IF}) that satisfies the following limit near the domain of $t^{'} \in (0,t_c)$:
\begin{equation}
    \lim_{t \to t^{'}} IF_{merger}(t) = IF^{chirp}_{linear}(\delta t, \Omega),
    \label{equ: chirp approx limit}
\end{equation}
where $\delta t$ is the time allowed for the linear chirp frequency to vary, and $\Omega$ is an arbitrary base frequency. Constrained by the sampling frequency of the input signal and the desired range of the frequency and chirp rate analysis, $\delta t$ and $\Omega$ have upper and lower limits when dealing with real detector signals\cite{2015aLIGO}.  \\
Given that a BBH merger GW waveform can be approximated by linear chirp signals with an appropriate base frequency, chirp rate, and time sampling interval, we propose that the merger signal within such a time interval be analyzed using linear chirp transform (LCT)\cite{alkishriwo_chaparro_2012, alkishriwo}. Instead of matching the detector signal $h(t)$ to a constant frequency signal model $e^{-i2 \pi\Omega t}$ in an FT, a linear chirp signal model $e^{-i2 \pi(\Omega t+\gamma t^2)}$ is used.
The continuous LCT is defined as:
\begin{equation}
H_{L}(\Omega,\gamma)=\int_{-\infty}^{\infty}{h(t)e^{-i2\pi(\Omega t+\gamma t^2)}dt},
\end{equation}
and the continuous inverse LCT is given by
\begin{equation}
h(t)=\int_{-\infty}^{\infty}{H_{L}(\Omega,\gamma)}e^{i2\pi(\Omega t+\gamma t^2)}d\Omega,
\end{equation}
where the optimal ranges of $\Omega$ and $\gamma$ are determined based on the target BBH merger population and GW detector specifications.

%%%%%%%%%%%%%%%%%%%%%%%%The Joint-Chirp-rate-Time-Frequency Transform%%%%%%%%%%%%%%%%%%%%%%
\subsection{The Joint-Chirp-Rate-Time-Frequency Transform}
\begin{figure}
    \includegraphics[width=0.48\textwidth]{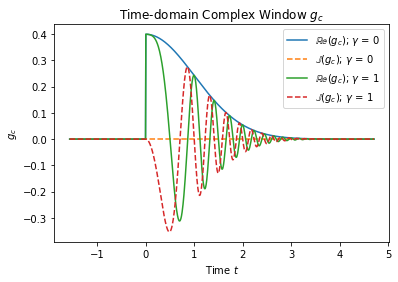} 
    \caption{Time-domain complex window function with $\Omega = 1 \;Hz$, $\Omega_0 = 0$, $\mu = 1$, $\tau = 0$, $\gamma =$ 0 and 1 $s^{-2}$, in the time range of $t \in [-1/2\pi, 3/2\pi]$. Solid lines represent real parts of the window and dashed lines represent imaginary parts of the window.}
    \label{fig:complex window in time}
\end{figure}

\begin{figure}
\includegraphics[width=0.48\textwidth]{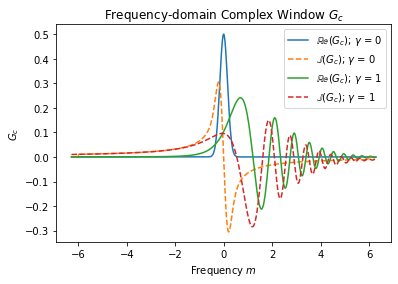} 
\caption{Frequency-domain complex window function with $\Omega = 1\;Hz$, $\Omega_0=0$, $\mu=1$, $\tau = 0$, $\gamma = 0$, and $1\; s^{-2}$ in the shifted frequency range of $\alpha \in [-2\pi, 2\pi]$. Solid lines represent real parts of the window and dashed lines represent imaginary parts of the window.}
\label{fig:complex window frequency domain}%
\end{figure}

We define the complex window function based JCTFT as
\begin{equation}
H_{J}(\gamma,\tau, \Omega)=\int_{-\infty}^{\infty}{h(t)g_{c}(\gamma, t-\tau, \Omega)e^{-i2\pi \Omega t}dt},
\label{equ:general JCTFT in time}
\end{equation}
where $h(t)$ is the input signal, $t$ is the input time, $\tau$ is the transformed time, $\gamma$ is the chirp rate, $\Omega$ is the frequency, and $g_{c}(\gamma, t-\tau, \Omega)$ is the complex window function. The complex window function is defined as follows:
\begin{equation}
    \begin{aligned}
    g_{c}(\gamma, t-\tau, \Omega) = 
    \frac{|\Omega_0+\mu\Omega|}{\sqrt{2\pi}}\frac{1+\frac{t-\tau}{|t-\tau|}}{2}\\
    e^{-(t-\tau)^2[\frac{(\Omega_0 +\mu\Omega)^2}{2}+i2\pi \gamma]},
    \label{equ:complex window in time}
    \end{aligned}
\end{equation}
where $\frac{|\Omega_0+\mu\Omega|}{\sqrt{2\pi}}$ is the partial normalization factor that ignores the contribution of the imaginary component, and $\Omega_0 +\mu\Omega$ in the exponent acts as a frequency-varying standard deviation control of the window function width. $\Omega_0$ is an arbitrary base frequency parameter and $\mu$ is the frequency scaling parameter, both of which can be determined experimentally. This frequency-varying window standard deviation control technique allows for the dynamic self-adjustment of the window width for improved resolution\cite{stockwell_mansinha_lowe_1996}. A set of these complex window functions is shown in Figure (\ref{fig:complex window in time}), where $\Omega=1 \;Hz$, $\Omega_0=0$, $\mu=1$, and the chirp rate $\gamma$ is equal to 0 and 1 $s^{-2}$. As stated by equation (\ref{equ:complex window in time}), the complex-valued window function is a typical Gaussian function with the frequency shift component of a linear chirp signal absorbed into itself.
\\
To make use of this frequency-related window function standard deviation control in equation (\ref{equ:complex window in time}), it is necessary to have prior knowledge of the input signal frequency distribution in time, which is not always possible. Fortunately, this problem can be solved efficiently by converting the JCTFT windowing operation from the time domain to the frequency domain. Instead of sliding the complex window function in time and transforming the windowed signal, we first obtain the frequency distribution of the input signal via an FT and then slide a complex window function along the frequency axis of the signal. It is easy to show that the FT of a window function $g_c(t-\tau)$ with a time shift $\tau$ is
\begin{equation}
    \mathcal{F}\{g_c(t-\tau)\} = \mathcal{F}\{g_c(t)\}e^{-i2\pi \Omega\tau},
\end{equation}
where $\mathcal{F}$ is the FT operator and all other variables are treated as coefficients for a particular window. The JCTFT can be represented as follows with the time shift factored out explicitly:
\begin{equation}
    \begin{aligned}
    H_{J}(\gamma, \tau, \Omega)= \mathcal{F}\{f(t)g_c(t-\tau, \gamma, \Omega)\} \\
    = \mathcal{F}\{f(t)\} \ast \mathcal{F}\{g_c(t-\tau, \gamma, \Omega)\} \\
    = \mathcal{F}\{f(t)\} \ast \mathcal{F}\{g_c(t, \gamma, \Omega)\} e^{-i2\pi \Omega\tau},
    \end{aligned}
\end{equation}
where the convolution theorem is used to convert the FT of multiplication in the time domain to convolution in the frequency domain, which is represented by the convolution operator $*$. Finally, an alternative frequency domain definition of the JCTFT is
\begin{equation}
H_{J}(\gamma, \tau, \Omega)=\int_{-\infty}^{\infty} H(\Omega+\alpha) G_{c}(\gamma, \Omega, \alpha) e^{i2\pi \alpha \tau} d\alpha,
\label{equ:frequency JCTFT simple}
\end{equation}
where $\alpha$ is the frequency shift parameter introduced by the definition of convolution, $H(\Omega+\alpha)$ is the FT of the original input signal on the shifted frequency scale, and $G_c(\gamma, \Omega, \alpha)$ is the FT of the complex window function $g_c(\gamma, t, \Omega)$ without time shift $\tau$. 
\begin{equation}
    G_c(\gamma, \Omega, \alpha)=\frac{|\Omega_0 +\mu\Omega|}{\sqrt{2\pi}}\{e^{-\pi^2 \alpha^2/z}(\frac{1}{2}\sqrt{\frac{\pi}{z}}[1 - erf(\frac{i\pi \alpha}{\sqrt{z}})])\},
    \label{equ:complex window in frequency}
\end{equation}
where $z = a+ib$, $a = \frac{(\Omega_0 +\mu\Omega)^2}{2}$, and $b = 2\pi\gamma$. The alternative JCTFT definition equation (\ref{equ:frequency JCTFT simple}) takes a form similar to the S transform\cite{stockwell_mansinha_lowe_1996}, with the difference being that $G_c$ is complex and contains a frequency-shift term. Figure (\ref{fig:complex window frequency domain}) shows a set of complex window functions given by equation (\ref{equ:complex window in frequency}) with $\Omega = 1\;Hz$, $\Omega_0=0$, $\mu=1$, $\tau = 0$, chirp rate $\gamma = $ 0, and 1 $s^{-2}$. 
\\
The convolution shift parameter $\alpha$ vanishes after the integration. According to the definition of convolution, this frequency shift should occur in the opposite direction such as $H(\Omega - \alpha)$ and $G_c(\gamma,\Omega, -\alpha)$. However, by invoking the correlation theorem after convolution, this frequency shift can be interpreted as $H(\Omega - (-\alpha))$ and $G_c(\gamma, \Omega, -(-\alpha))$. This step is important for discrete JCTFT computations as it transforms a convolution operation to an inverse FT, which can be computed efficiently using the inverse fast FT (IFFT) algorithm on the complexity scale of $n\;log(n)$\cite{10.2307/2003354}. \\
The discrete JCTFT referencing equations (\ref{equ:frequency JCTFT simple}) and (\ref{equ:complex window in frequency}) is given by
\begin{equation}
    H_J[l, m, k]= \frac{1}{N}\sum_{j=0}^{N-1}H[k+j]G_c[l,k,j]e^{\frac{i2\pi}{N}jm},
    \label{equ:discrete JCTFT}
\end{equation}
where $l$ is the discrete chirp rate number, $m$ is the discrete transformed time, $k$ is the discrete frequency, $j$ is the discrete shifted frequency, and $N$ is the total number of the original discrete time. Here, the discrete time and frequency share the same length $N$.
By the convention established earlier, $H[k+j]$ represents the discrete FT (DFT) of the input signal $h[n]$,
\begin{equation}
    H[k+j] = \sum_{n=0}^{N-1}h[n]e^{-\frac{i2\pi}{N}(k+j)n}.
\end{equation}
$G_c[l,k,j]$ represents the DFT of the discrete complex window function in equation (\ref{equ:complex window in frequency}).
\begin{equation}
    G_c[l, k, j]=\frac{|k_0 +\mu k|}{\sqrt{2\pi}}\{e^{-\pi^2 j^2/v}\frac{1}{2}\sqrt{\frac{\pi}{v}}[1 - erf(\frac{i\pi j}{\sqrt{v}})]\},
    \label{equ:discrete complex window in frequency}
\end{equation}
where $v = c+id$, $c = \frac{(k_0 +\mu k)^2}{2}$, and $d = 2\pi Cl$. The discrete chirp rate is represented by the product of a constant scaling coefficient $C$ and the discrete chirp rate number $l$. $k_0$ is the discrete equivalent of the base frequency $\Omega_0$. \\
Similar to STFT, which uses a real window function $g(t-\tau)$, JCTFT is invertible. More precisely, the solution of JCTFT corresponding to any chirp rate $\gamma$ alone contains the full information required for an inverse transform. When the chirp rate $\gamma = 0$, the time-frequency decomposition is equivalent to an S transform, and the inverse transform follows the inverse S transform algorithm\cite{stockwell_mansinha_lowe_1996}. For JCTFT solutions with non-zero $\gamma$ values, corrections to the frequency shifts introduced by the linear chirp component are needed. In general, the partially normalized complex window function in equation (\ref{equ:complex window in time}) has
\begin{equation}
    \int_{-\infty}^{\infty}g_c(\gamma, t-\tau, \Omega) dt = \sqrt{\frac{(\Omega_0 +\mu\Omega)^2}{4(\Omega_0 +\mu\Omega)^2+i16\pi\gamma}}.
    \label{equ:time complex window integral}
\end{equation}
An immediate consequence of equation (\ref{equ:time complex window integral}) is
\begin{equation}
    A \int_{-\infty}^{\infty}g_c(\gamma, t-\tau, \Omega) dt = 1,
\end{equation}
with $A=(\frac{(\Omega_0 +\mu\Omega)^2}{4(\Omega_0 +\mu\Omega)^2+i16\pi\gamma})^{-1/2}$. When combined with the inverse FT, the properly normalized inverse JCTFT (IJCTFT) is given by
\begin{equation}
    h(t) = \int_{-\infty}^{\infty}\int_{-\infty}^{\infty}A\cdot H_{\gamma}(\tau, \Omega) d\tau \; e^{i2\pi\Omega t}d\Omega,
\end{equation}
where $H_{\gamma}$ is the $\gamma$ portion of the JCTFT result. The discrete IJCTFT is obvious:
\begin{equation}
    h[n] = \sum_{j=0}^{N-1}\frac{1}{N}\sum_{m=0}^{N-1}A\cdot H_{\gamma}[m,j]e^{i2\pi jn}.
    \label{equ:discrete IJCTFT}
\end{equation}
In practice, other approaches also exist to invert a discrete JCTFT, some of which can be more efficient than equation (\ref{equ:discrete IJCTFT}). For example, the effect of a window function can be reversed for a time interval simply by applying the inverse of the particular function. Zero-division errors must be avoided because the complex windows $g_c(\gamma, t-\tau, \Omega)$ and $G_c(\gamma, \Omega, \alpha)$ may have multiple zero-crossings. 

\subsection{Chirp Rate Enhanced Time-Frequency Spectrogram}
The result of a JCTFT has three dimensions: chirp rate, time, and frequency. In many cases, having two-dimensional representations of a higher-dimensional dataset simplifies the analysis, given that these lower-dimensional representations capture a sufficient amount of detail of the original data. With this assumption, in this section, we define an operation based on the result of a JCTFT for time-frequency spectrogram generation. The spectrogram generation operation is achieved by taking an orthogonal projection of JCTFT along the chirp rate axis:
\begin{equation}
    S_J(\Omega,\tau)=\int_{-L}^{L} H_J(\gamma, \tau, \Omega) d\gamma,
    \label{equ:spectrogram from JCTFT}
\end{equation}
where the chirp rate $\gamma$ is assumed to have a range of $[-L, L]$ for $L \in \mathbb{R}$. The discrete operation follows 
\begin{equation}
    S_J(m, j) = \sum_{l=-l}^{l} H_J[l, m, j].
\end{equation}
where $l = L/C$.
The integral in equation (\ref{equ:spectrogram from JCTFT}) suggests that the individual chirp rate distribution is lost during the process, making this operation non-invertible. Custom window functions tailored for generating desired time-frequency spectrograms can be further developed to enhance the complex window function. For instance, an additional window function for removing weaker frequency components can be defined as
\begin{equation}
    g_{freq}(\Omega) = e^{-\frac{(\Omega-\Omega_{peak})^2}{2\sigma^2}},
    \label{equ: special window Gaussian}
\end{equation}
where $\Omega_{peak}$ is the peak frequency response position of the signal for a given chirp rate, and $\sigma$ is an independent standard deviation control parameter. The quality of a custom window function largely dictates the quality of the generated spectrogram. However, we do not discuss window design in detail here. It should be mentioned that the chirp-rate-enhanced time-frequency spectrogram must not be treated as an accurate time-frequency representation of the input signal as the frequency content of the input signal has been modified by the complex window function. JCTFT is designed with easier chirp signal detection in mind rather than as an alternative to traditional time-frequency decomposition methods.\\
Similarly, replacing the chirp rate integral with a frequency integral produces the chirp rate variant of the projection operation:
\begin{equation}
    C_J(\gamma,\tau)=\int_{-\infty}^{\infty} H_J(\gamma, \tau, \Omega) d\Omega.
    \label{equ:chirp-time from JCTFT}
\end{equation}
Because of the typical wide signal representation in the chirp-rate-frequency domain along the chirp rate axis, an efficient custom window design for equation (\ref{equ:chirp-time from JCTFT}) is particularly difficult.

%%%%%%%%%%%%%%%%%%%%%%%%The Joint-Chirp-rate-Time-Frequency Transform%%%%%%%%%%%%%%%%%%%%%%
\section{Results}
\subsection{Three-dimensional JCTFT Result}
\begin{figure*}[!ht]
    \centering
    \includegraphics[width=0.8\textwidth]{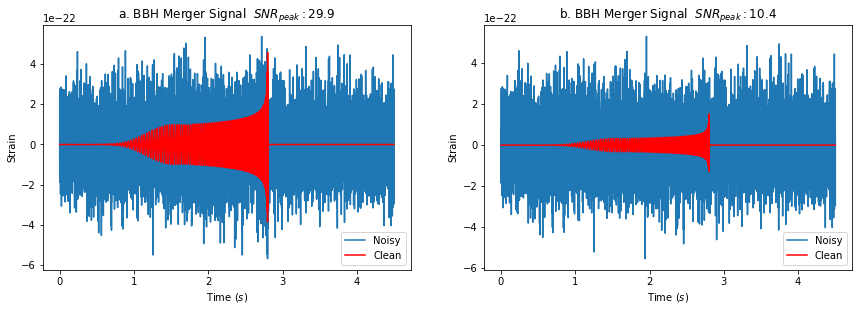}
    \caption{ (a) Signal $h^{29.9}_{SNR}(t)$: Simulated high SNR BBH merger GW signal strain $h^{600}_{Mpc}(t)$ with aLIGO detector characteristic noise. (b) Signal $h^{10.4}_{SNR}(t)$: Simulated low SNR BBH merger GW signal strain $h^{1800}_{Mpc}(t)$ with aLIGO detector characteristic noise. Noisy signals $h^{29.9}_{SNR}(t)$ and $h^{10.4}_{SNR}(t)$ are plotted in blue. Merger signals are plotted in red.}
    \label{fig:noisy merger}%
\end{figure*}

\begin{figure*}[!ht]
    \centering
    \includegraphics[width=0.725\textwidth]{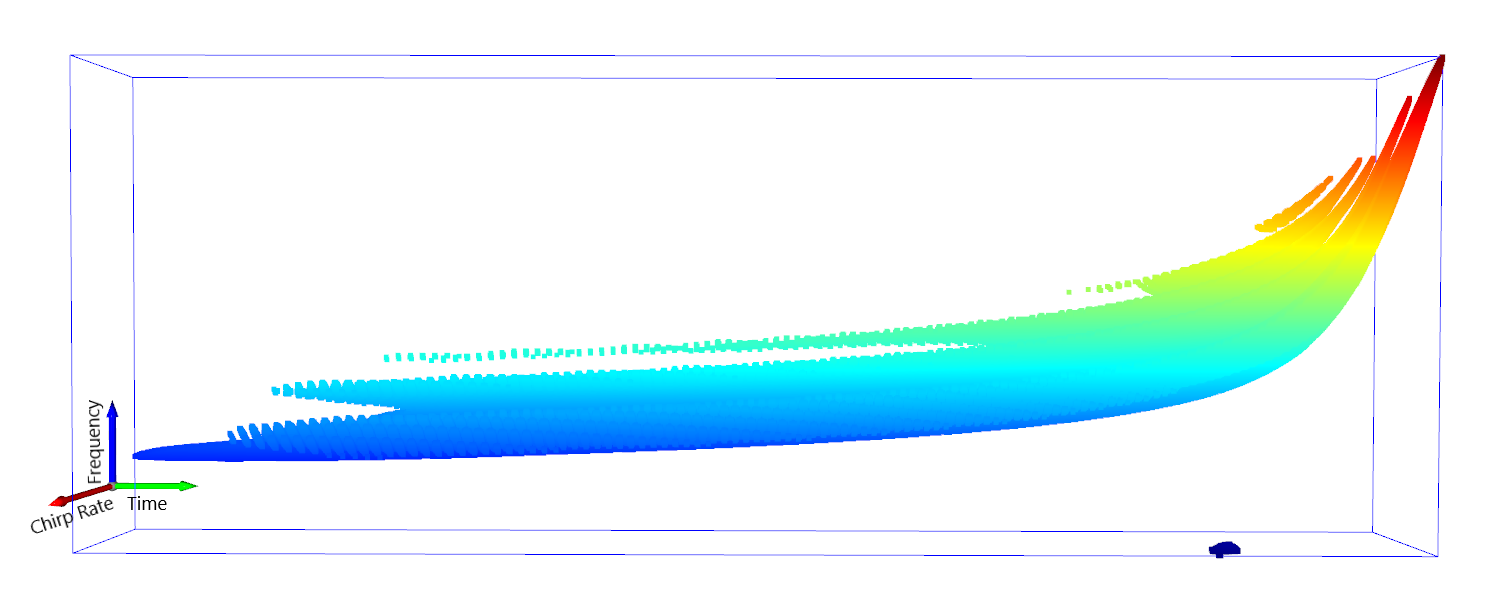} 
    \caption{Processed JCTFT result of merger signal $h^{600}_{Mpc}(t)$ plotted in three-dimensional space. The green, blue, and red axes indicate time, frequency, and chirp rate. Red voxels represent higher frequencies, and blue voxels represent lower frequencies.}
    \label{fig:JCTFT-time-freq}%
\end{figure*}

\begin{figure*}[!ht]
    \centering
    \includegraphics[width=0.725\textwidth]{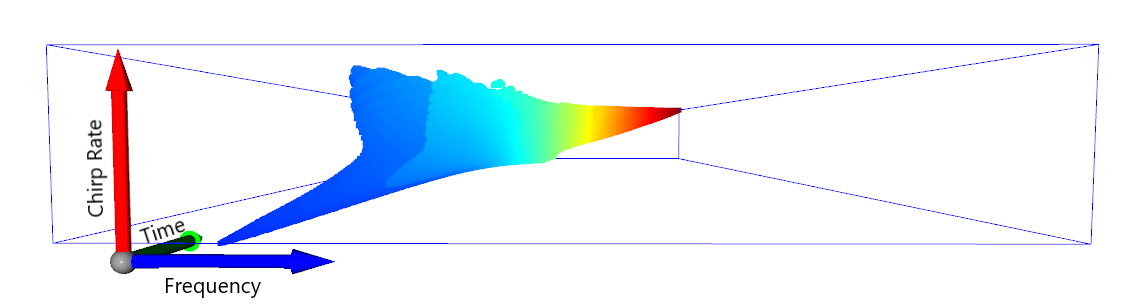} 
    \caption{Processed JCTFT result of merger signal $h^{600}_{Mpc}(t)$ plotted in three-dimensional space. The green, blue, and red axes indicate time, frequency, and chirp rate. Red voxels represent higher frequencies, and blue voxels represent lower frequencies.}
    \label{fig:JCTFT-chirp-freq}%
\end{figure*}

Let $h^{600}_{Mpc}(t)$ and $h^{1800}_{Mpc}(t)$ be BBH merger GW signals from systems with $m_1 = 15M_{\odot}$ and $m_2 = 10 M_{\odot}$ at distances of 600 Mpc and 1800 Mpc generated using the frequency-domain phenomenological model IMRPhenomD\cite{PhysRevD.93.044007}. Figure (\ref{fig:noisy merger}) shows the merger signals with simulated advanced LIGO (aLIGO) noise. The clean strain $h^{600}_{Mpc}(t)$ in red in Figure (\ref{fig:noisy merger}a) has amplitudes around the order of $10^{-22}$ with a peak amplitude of approximately $4\cdot10^{-22}$. The clean strain $h^{1800}_{Mpc}(t)$ in Figure (\ref{fig:noisy merger}b) has a peak amplitude of approximately $1.7\cdot10^{-22}$. Parameters in Table (\ref{table:1}) were used to compute the JCTFT.
\begin{table}[ht]
\centering
\begin{tabular}{||c c||} 
\hline
Parameter & Value \\ [0.5ex] 
\hline\hline
Input Sample Frequency ($s^{-1}$)    & \;\;\;\;\;\;\;2048 \;\;\;\;\;\; \\    
JCTFT Sample Frequency $(s^{-1})$ & 600\\
$\Omega$ $(s^{-1})$            & 30 - 300  \\
$\Omega_0^*$ $(s^{-1})$        & 25\\
$\mu^*$               & 0.0075 \\
$\gamma^*$ $(s^{-2})$     & 0 - 2400\\\
$C^*$ $(s^{-2})$   & 60\\
$l^*$    & 0 - 40\\
$\sigma^* (s^{-1})$     & 6\\
[1ex] 
\hline
\end{tabular}
\caption{JCTFT parameter values or ranges used for generating the results in this paper unless specified otherwise. The JCTFT sampling frequency and $\Omega$ are determined based on detector specifications\cite{2015aLIGO}. Parameters labeled by $^*$ are estimated based on determined values using equations (\ref{equ: IF merger ^(-8/3)}), (\ref{equ: chirp approx limit}), and (\ref{equ:general JCTFT in time}). Different combinations may produce better or worse results.}
\label{table:1}
\end{table}
Figures (\ref{fig:JCTFT-time-freq}) and (\ref{fig:JCTFT-chirp-freq}) show three-dimensional JCTFT voxels with the top $1\%$ highest values for visualization purposes. Voxels represent signal presence near the neighborhood. Figure (\ref{fig:JCTFT-time-freq}) shows the processed JCTFT result of $h^{600}_{Mpc}(t)$ viewed from the time-frequency front. Reddish colors represent higher frequencies and bluish colors represent lower frequencies. As time progresses in the positive direction indicated by the green axis in the lower left corner of Figure (\ref{fig:JCTFT-time-freq}), the main frequency feature also climbs higher. This is visible from the redder voxels that reach higher values in the positive frequency direction labeled by the blue axis.
 Figure (\ref{fig:JCTFT-chirp-freq}) shows the JCTFT result viewed from the chirp-rate-frequency front. The positive chirp rate direction is indicated by the red axis and the positive frequency direction is indicated by the blue axis in the lower-left corner of Figure (\ref{fig:JCTFT-chirp-freq}). The lowest frequency components located near the lower-left corner have lower chirp rates. As frequency increases in the positive blue axis direction, the signal chirp rate first increases, and then develops branches visible in both Figure (\ref{fig:JCTFT-time-freq}) and Figure (\ref{fig:JCTFT-chirp-freq}). The higher frequency components on the right side of Figure (\ref{fig:JCTFT-chirp-freq}) show redder colors, which indicate that higher values and higher frequency voxels have higher chirp rates.

%%%%%%%%%%%%%%%%%%%%%%%%%%%%% Time Frequency Spectrogram %%%%%%%%%%%%%%%%%%%%%%%%
\subsection{Time-Frequency Spectrogram}
\begin{figure*}[!ht]
    \centering
    \includegraphics[width=0.8\textwidth]{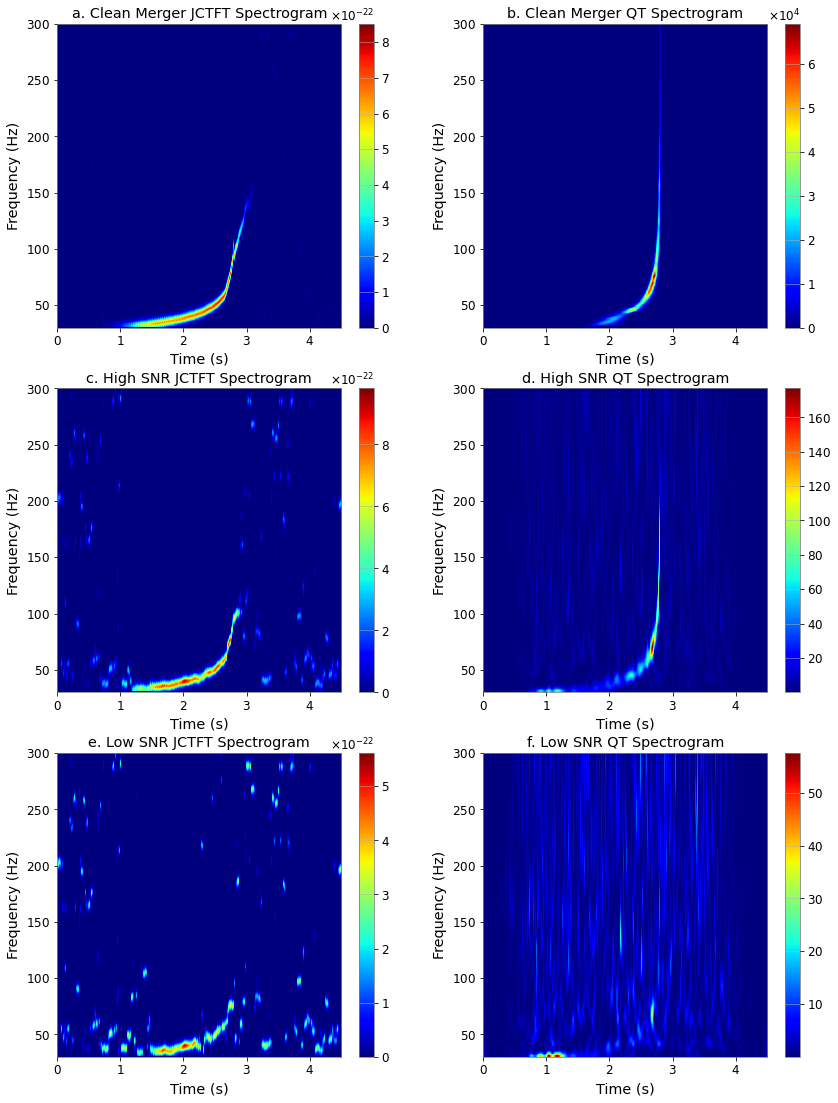}
    \caption{(a) JCTFT spectrogram of the clean merger signal $h^{600}_{Mpc}(t)$. (b) QT spectrogram of the clean merger signal $h^{600}_{Mpc}(t)$. (c) JCTFT spectrogram of $h^{29.9}_{SNR}(t)$. (d) QT spectrogram of $h^{29.9}_{SNR}(t)$. (e) JCTFT spectrogram of $h^{10.4}_{SNR}(t)$. (f) QT spectrogram of $h^{10.4}_{SNR}(t)$.}
    \label{fig:3*2 Spectrogram Collection}
\end{figure*}
The time-frequency projection of a JCTFT by definition is comparable to that of a traditional transform with a low-to-moderate amount of noise in the input signal and a sufficient time interval sampling rate. We compare the JCTFT time-frequency spectrograms of simulated merger signals with three different SNR levels to the spectrograms generated by Q-Transform. Transformations in the QT family are widely used in audio analysis and signal processing because their logarithmically scaled, and frequency-varying window functions are well-suited for signals that span across large frequency scales \cite{brown_1991}. Additional controls over  window functions in the QT are introduced by a quality factor Q where larger Q dilates windows across all frequencies and smaller Q does the opposite. For this reason, a QT with dynamically determined Q-factors is one of the most commonly used transformation techniques in GW merger signal visualization and analysis. The following QT spectrograms were generated using the Q-factor optimization and QT algorithms in GWDetChar \cite{gwdetchar} and GWpy \cite{gwpy}, the algorithms used in the official LIGO detector characterization \cite{Davis_2021}. The JCTFT spectrogram values represent the combined matching amplitudes and the QT spectrograms represent the normalized energy. Figure (\ref{fig:3*2 Spectrogram Collection}a) shows the JCTFT time-frequency projection of the clean merger signal $h^{600}_{SNR}(t)$, which we refer to as spectrogram for simplicity. Figure (\ref{fig:3*2 Spectrogram Collection}b) shows the QT spectrogram of the slightly whitened clean merger signal $h^{600}_{Mpc}(t)$. The whitening was applied for optimal QT performance. Both spectrograms display the peak merger frequency near $2.5\;s$ to $2.70\; s$. The JCTFT spectrogram contains a more prominent inspiral stage signal between 1 and 2 seconds when compared to the QT spectrogram. The QT spectrogram peak merger frequency component extends to approximately $300 Hz$, and the JCTFT spectrogram merger frequency reaches only approximately $150 Hz$. \\
Substituting the high SNR simulated noisy strain $h^{29.9}_{SNR}(t)$ into equation (\ref{equ:spectrogram from JCTFT}) with the additional frequency domain window function equation (\ref{equ: special window Gaussian}) yields the JCTFT spectrogram shown in Figure (\ref{fig:3*2 Spectrogram Collection}c). The QT spectrogram of $h^{29.9}_{SNR}(t)$ is shown in Figure (\ref{fig:3*2 Spectrogram Collection}d). Both the JCTFT and the QT spectrograms show reduced peak merger frequency as a result of the increased amount of simulated background noise. The faint inspiral stage signal is visually more pronounced when comparing Figure  (\ref{fig:3*2 Spectrogram Collection}c) to (\ref{fig:3*2 Spectrogram Collection}d). \\
Substituting $h^{10.4}_{SNR}(t)$ into the JCTFT time-frequency projection and the QT yields the JCTFT (Figure \ref{fig:3*2 Spectrogram Collection}e) and QT (Figure \ref{fig:3*2 Spectrogram Collection}f) spectrograms of the low SNR signal. Both $h^{10.4}_{SNR}(t)$ JCTFT and QT spectrograms are of much worse quality than $h^{29.9}_{SNR}(t)$ and $h^{600}_{Mpc}(t)$ spectrograms (for the purpose of merger detection). The lower frequency inspiral signal is less consistent, with significant amounts of discontinuity caused by the simulated aLIGO noise. 

%%%%%%%%%%%%%%%%%%%%%%%%%%%%%% Classification Results %%%%%%%%%%%%%%%%%%%%%%%%%%%%
\subsection{Spectrogram Classification Accuracy}
\begin{figure}
    \centering
    \includegraphics[width=0.48\textwidth]{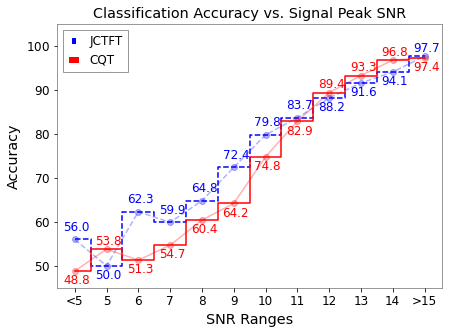} 
    \caption{InceptionV3 network classification accuracies of the SNR-categorized testing dataset. The two networks were trained separately using the JCTFT and QT-generated spectrograms.}
    \label{fig:snr accuracy}%
\end{figure}
\begin{figure*}
    \centering
    \includegraphics[width=1\textwidth]{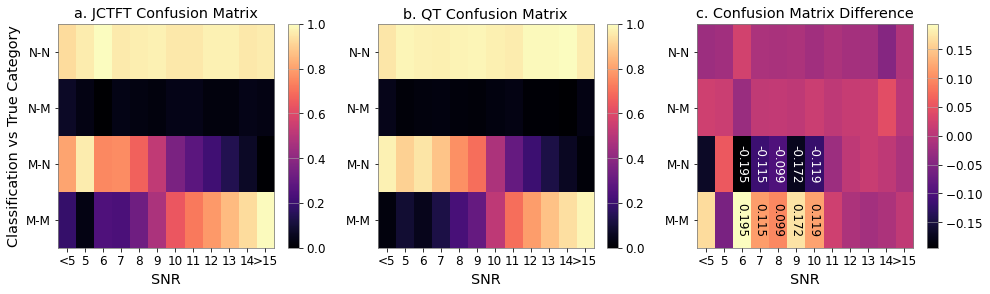} 
    \caption{The confusion matrices of (a) the JCTFT-trained, and (b) the QT-trained InceptionV3 classification network evaluated using the SNR-categorized testing dataset. (c) is the difference between the JCTFT and QT matrices. Selected difference values are labeled.}
    \label{fig:snr confusion}%
\end{figure*}

The InceptionV3 is a deep convolutional neural network architecture that has demonstrated state-of-the-art image classification performance on the ImageNet classification dataset \cite{inceptionV3, imageNet}. We use an InceptionV3 network with pre-trained ImageNet weights as the base model for further training and evaluating the merger detection accuracy using JCTFT and QT spectrograms.
The training dataset is comprised of two subsets, M1200 and M2000. M1200 contains 12,500 simulated noisy merger waveforms from BBH systems between 600 Mpc and 1200 Mpc, and 12,500 aLIGO detector noise waveforms; M2000 contains 12,500 simulated noisy merger waveforms from systems between 1400 Mpc and 2000 Mpc, and 12,500 aLIGO detector noise waveforms. The merger waveforms were simulated for systems with combined pre-merger masses between $10M_{\odot}$ and $83M_{\odot}$. The merger waveforms were randomly injected into the 2-second, 3-second, and 4-second positions of the 6-second noise waveforms with an additional 0.5 seconds of random shifts. Table (\ref{table:A1}) contains details of the dataset specifications. \\
We train the two InceptionV3 networks using JCTFT and QT spectrograms separately in two stages, first with the M1200 and then with the M2000 subset. The JCTFT-trained network achieved 98.26\% accuracy for the M1200 dataset and 91.6\% accuracy for the M2000 dataset. The QT-trained network achieved 97.3\% and 89.02\% respectively.
We evaluate the performance of both InceptionV3 networks using a set of 4000 newly generated waveforms, where half of which are noisy merger waveforms generated following the same specifications as in Table (\ref{table:A1}). The classification performance was evaluated in 12 merger signal SNR categories: less than 5, 5 – 14 in intervals of 1, and greater than 15. Figure (\ref{fig:snr accuracy}) shows the classification accuracy vs. signal SNR ranges. Between SNR 6 to 10, the InceptionV3 model trained on JCTFT spectrograms performed consistently better than the same model trained with QT spectrograms. The JCTFT-trained network saw an 11\% improvement in classification accuracy at SNR 6. Cases of 7, 8, 9, and 10 each saw improvements of 5.2\%, 4.4\%, 8.2\%, and 5\%. The QT-trained network performed slightly better with SNR 12, 13, and 14. For all signals with SNR higher than 15, both networks performed equally well with minimal difference.\\
Figures (\ref{fig:snr confusion}a) and (\ref{fig:snr confusion}b) show the network confusion matrices of both networks evaluated using the SNR-categorized dataset. Merger-Merger (MM) and Noise-Noise (NN) represent the percentage of merger signals being correctly classified as mergers or noises. Merger-Noise (MN) and Noise-Merger (NM) represent the percentage of mergers or noises being misclassified as opposites. Both the JCTFT and QT-trained InceptionV3 networks were able to classify the noise signals to very high accuracies. Higher SNR corresponds to higher MM accuracy. Lower SNR corresponds to a higher NM rate.  Figure (\ref{fig:snr confusion}c) demonstrates the difference in the confusion matrices of the JCTFT-trained network and the QT-trained network.  The JCTFT-trained network has consistently higher MM accuracy and lower MN error rate than the QT-trained network. For example, the JCTFT-trained network saw a 19.5\% improvement in MM accuracy at SNR 6. Cases with SNR of 7, 8, 9, and 10 each showed a growth of 11.5\%, 9.9\%, 17.2\%, and 11.9\% in MM accuracy.
%%%%%%%%%%%%%%%%%%%%%%%%%%%%%%%%%%% Discussions %%%%%%%%%%%%%%%%%%%%%%%%%%%%%%%%%%%
\section{Discussion}
\subsection{Three-dimensional JCTFT}
A chirp signal can be traced in the additional chirp rate dimension of the three-dimensional JCTFT solution. In contrast to having only the time-frequency spectrogram, JCTFT provides higher matching amplitudes of a signal within a short time interval when the chirp signal model matches the input chirp signal well. This is demonstrated in the $h^{600}_{Mpc}(t)$ JCTFT solution viewed from the chirp-rate-frequency front in Figure (\ref{fig:JCTFT-chirp-freq}). All voxels with values below the top 99 percentile were removed from the result. The zero chirp rate time-frequency plane resembles the traditional time-frequency spectrogram of the merger signal $h(t)$, which only shows lower frequency components of the signal. This is due to the lower frequency inspiral signals having lower chirp rates. Higher-frequency merger signals have higher chirp rates and, as a result, the best matching merger stage voxels have higher frequencies and higher chirp rates. This technique accumulates the chirp signal matching amplitude with the inclusion of chirp rates and produces clearer frequency domain peaks than traditional transform methods. In Figure (\ref{fig:JCTFT-chirp-freq}), the abundance of high chirp rate artifacts characterizes the JCTFT result. By comparing the time-frequency plane of different non-zero chirp rate values to the zero chirp rate spectrogram and analyzing the JCTFT definition, we conclude that these voxels are caused by partially matched linear chirp waveforms. The starting frequencies of these partially matched waveforms are different from the actual starting frequency of the input chirp signal. However, when paired with the complex window function, the IF of these waveforms increases or decreases to a degree that matches a significant portion but not all of the input chirp signal, generating artifacts in the form of branches in the three-dimensional JCTFT result. Fortunately, these artifacts can easily be removed by masking functions. 
Tracing high matching amplitudes in the chirp rate, time, and frequency domains provides a new way of identifying the chirp signal and paves the ground for new three-dimensional analysis applications. 

\subsection{JCTFT and QT Spectrograms}
Valuable information about the input signal can be extracted from the two-dimensional JCTFT time and frequency projections. The lower frequency bound of the clean merger signal $h^{600}_{Mpc}(t)$ was set to $30 \;Hz$. The aLIGO noise model also has a lower frequency cut-off of approximately $30 \;Hz$. 
The QT spectrogram in Figure (\ref{fig:3*2 Spectrogram Collection}f) displays a cluster of peak normalized energy centered around 1 second in time and 30 Hz in frequency. This peak normalized energy value is so high relative to all other energy tiles in the QT spectrogram that the main inspiral stage signal of the merger after the first second can only be seen vividly in the background. On the contrary, the combined matching amplitudes of the JCTFT remained in the order of $10^{-22}$ for all three different SNR levels. In addition, the peak amplitude positions in time and frequency remained relatively stable. For lower SNR merger signals, the dynamic Q-factor optimization and energy normalization processes in the QT, originally designed to improve the spectrogram quality, becomes a disadvantage.
The JCTFT time-frequency spectrograms might perform better than other transformation techniques for some specialized purposes. However, chirp-rate-enhanced JCTFT spectrograms should not be treated like traditional spectrograms generated using methods such as the STFT, ST, QT, or most wavelet transforms. For each small interval in time, the JCTFT result contains implicit frequency drifts that do not show up in the time-frequency projection. Additional frequency features might be present due to the frequency shifts in the JCTFT, for example, the peak merger frequency feature between 2.7 and 3 seconds in Figure (\ref{fig:3*2 Spectrogram Collection}a). We would not expect features like this in traditional time-frequency spectrograms. A clean linear chirp signal from $\omega_0$ to $\omega_1$ with a constant chirp rate $\gamma_0$ triggers narrow and well-isolated frequency peaks at the starting frequency of each small time interval and chirp rate $\gamma_0$. In contrast, traditional decomposition methods show continuously varying frequency distributions across $\omega_0$ to $\omega_1$ but with lower peak amplitudes. The JCTFT spectrogram prohibits accurate representation of the signal in the time-frequency domain alone unless corrections are made using the chirp rate distribution information. In cases with high time-sampling rates or low chirp rates, one can treat the JCTFT time-frequency distribution as a traditional spectrogram, keeping the above differences in mind. 

%%%%%%%%%%%%%%%%%%%%%%%%%%%%%% Classification Accuracy %%%%%%%%%%%%%%%%%%%%%%%%%%%%
\subsection{Neural Network Spectrogram Classification Accuracy}
Figure (\ref{fig:snr accuracy}) shows consistently better performances of the JCTFT-trained model for merger signals with SNRs between 6 and 10. The QT-trained network performed slightly better than that of the JCTFT for datasets with SNR of 12, 13, and 14. Combined with the visual comparison of the spectrograms in Figure (\ref{fig:3*2 Spectrogram Collection}), we conclude the QT of high and moderate SNR cases provides spectrograms that are easier for image classification networks like the InceptionV3. This is because the Q-factor optimization and energy normalization work in favor of isolating the merger signal from background noise. On the contrary, when the SNR is low, the Q-factor optimization and the energy normalization introduce more uncertainty as to whether the transformed peak normalized energy is the peak of desired merger signals. The JCTFT complex chirp window function and the summation technique for time-spectrogram generation were designed with preserving more information about the desired chirp signal in mind rather than an alternative transformation technique that performs better for all cases. The main disadvantages of the QT in low SNR merger spectrogram generation tasks are the difficulty in determining the optimal quality factor. The normalized energy is also a metric used in determining the optimal Q-factor during the factor optimization process which could potentially lead to the choice of a less ideal Q-factor due to the influence of heavy non-stationary background noise. This could produce peak normalized energy that is several magnitudes higher than the rest of the signal, overpowering the useful information about the hidden merger signal that ANN requires for making informed predictions. The main objective of the JCTFT is to optimize its window function while preserving as much of the original merger information as possible. As a consequence, even when the generated spectrogram is littered with heavy background noise, there could still be enough information for the classification network to predict whether a merger signal is hidden in the spectrogram.

%%%%%%%%%%%%%%%%%%%%%%%%%%%%%% Cases of Application %%%%%%%%%%%%%%%%%%%%%%%%%%%%
\subsection{Cases of Application}
There are multiple ways to define the JCTFT. Each definition leads to similar results but with slightly different approaches to the signal decomposition problem. JCTFT works well when the input signal within each short time interval is well-matched by the linear chirp model. Alternatively, in our definition, equation (\ref{equ:general JCTFT in time}), the dominant frequency component in the windowed interval triggers a narrow and localized response in the Fourier space. This idea is not new in the field of BBH merger GW signal detection and other signal detection algorithms\cite{Klimenko_2016, gwpy}. It uses the complex window function to optimize the peak frequency response in a signal voxel in three-dimensional space or signal tile in two-dimensional projection.\\
The BBH merger GW signal is unique in that both the frequency and chirp rate span across large scales in a highly non-linear manner. Non-linear chirp rate scales should be used in discrete implementations of the JCTFT to efficiently cover the chirp rate space. In addition to chirp signal detection, JCTFT can be used for chirp rate and frequency estimations. The JCTFT produces consecutive peaks in the chirp-rate-frequency domain when a clear chirp signal is present. The chirp rate and frequency values for each time interval are the linear approximation of a section of the input signal.
\\
The main purpose of the JCTFT and JCTFT chirp rate enhanced spectrogram is to make identifying chirp signals from background noise easier rather than as an alternative method to traditional transformations. This objective is achieved by testing how the frequency-domain representation of a signal changes with different chirp rates. JCTFT enables new data analysis techniques by breaking down a time-series signal further than traditional time-frequency decomposition methods. The added chirp rate dimension opens up signal searches and analysis to three-dimensional spatial techniques and different ANN architectures. Examples include three-dimensional nearest neighbor search, correlation analysis, and even finding more efficient manifolds for two-dimensional signal representation with the help of machine learning algorithms\cite{arxiv.1802.03426}.

\subsection{Potential Improvements}
JCTFT is not a perfect solution for representing non-stationary signals, especially when faced with highly non-linear input signals and heavy background noise. The current GW detector signals fall into this category. Taking more discrete steps to linearly approximate a non-linear chirp signal may produce more accurate approximations, however, this process also makes the matched result prone to the influence of noise. A longer sampling duration accumulates the desired input signal, and the background noise can be analyzed by known noise models, but such an analysis requires a more complicated signal model design. The JCTFT is not restricted from using highly non-linear signal models, and the complex window function may take different forms. Therefore, it is necessary to balance the generality, effectiveness, and complexity of this method in practice.
Correcting the frequency shift caused by the complex window and denoising the result of a JCTFT time-frequency spectrogram using the chirp rate information will allow some infrastructures developed for current spectrogram-based search methods to be transferred for the JCTFT with minor tuning. We are working on a more detailed analysis of the JCTFT characteristics and frequency shift correction in immediate follow-up studies.

\section{Conclusions}
The JCTFT decomposes time-series signals into chirp rate, time, and frequency, and establishes the relationship between the three quantities. The linear chirp approximation gives the JCTFT, and methods extended from it, the ability to model continuous chirp signals better than traditional time-frequency decomposition methods. These characteristics of the JCTFT pave the way for new three-dimensional chirp signal searches and analysis techniques, using either classic methods or machine learning algorithms. The time-frequency projection spectrogram of the JCTFT result resembles a regular time-frequency spectrogram generated by traditional methods with appropriate choices of transform parameters. The use of additional chirp rate information and frequency window functions provides the knowledge required for generating chirp-enhanced spectrograms which makes JCTFT well-suited for chirp signal detection and analysis. InceptionV3 image classification network trained with JCTFT-generated spectrograms was able to achieve an average of 14\% better performance for simulated BBH merger signals with SNR 6-10. The JCTFT is a general signal decomposition technique that is applicable to a wide variety of signals, and can be applied to current generation ground-based GW detectors and future generation detectors, including space-based detectors, such as the Laser-Interferometer Space Antenna for long-duration and small chirp rate inspiral stage signal detection.\\
We propose quantitative studies of the effect of JCTFT in BBH and BNS merger GW signal detection pipelines and compare the results with pipelines using other techniques. Efficient non-templated searches powered by JCTFT processed data have the potential to reveal faint and distant GW sources with improved confidence.

\section*{Acknowledgements}
S.R. Valluri acknowledges partial support of this work through his NSERC Discovery Grant. X. Li and S.R. Valluri would like to thank Dr. Ramit Dey for his insightful suggestions throughout this project. We thank the authors of open-source Python libraries PyCBC, GWpy, SciPy, NumPy, Matplotlib, and Open3D\cite{Nitz:2017svb, gwpy, 2020SciPy-NMeth, harris2020array, Hunter:2007, open3d}, which have significantly accelerated signal generation, FFT calculation, and visualization processes.
The authors declare that they have no affiliations with or involvement in any organization or entity with any financial interest in the subject matter or materials discussed in this manuscript. 
\section*{Appendix}
\subsection*{A1. Noisy Merger Waveform Dataset Specifications}
\setcounter{table}{0}
\renewcommand{\thetable}{A\arabic{table}}
\begin{table}[ht]
\centering
\begin{tabular}{||c c||} 
 \hline
Parameters & Values \\[0.5ex] 
 \hline\hline
Sample Frequency ($s^{-1}$)    & 2048      \\
Lower-Frequency Cut-off  ($s^{-1}$) & 30   \\
Spectrogram Shape (pixels) & 256*256*3\\
Combined Pre-merger Mass ($M_{\odot}$) & 10 - 83 \\
Distances (kMpc) & 0.6 - 1.2 \\
                & 1.4 - 2\\
Z-axis Spin & [-1,1] \\
Spin Beta Distribution  &$\alpha = \beta = 0.125$\\
Merger Position ($s$)& 2, 3, or 4 $\pm$ 0.5\\
            & Normal Distribution\\[1ex] 
 \hline
 
\end{tabular}
\caption{Training and Testing Dataset Parameter Values.}
\label{table:A1}
\end{table}
\nocite{*}
\bibliographystyle{plainnat}
\bibliography{apssamp}% Produces the bibliography via BibTeX.

\end{document}